\documentclass[%
prl,
twocolumn,
superscriptaddress,
 amsmath,amssymb,
 aps,
]{revtex4-2}

\usepackage{graphicx}
\usepackage{dcolumn}
\usepackage{bm}
\usepackage[version=3]{mhchem}

\usepackage{tikz-cd}

\usepackage[normalem]{ulem}
\usepackage{todonotes}

\usepackage[output-decimal-marker={.},exponent-product=\cdot]{siunitx} 

\usepackage{hyperref}
\hypersetup{colorlinks=true,citecolor=blue,linkcolor=blue,filecolor=blue,urlcolor=blue,pdfstartview=FitH,pdfpagemode=UseNone}
\usepackage{tikz}

\newcommand{\R}{\mathbb{R}}
\newcommand{\C}{\mathbb{C}}

\newcommand{\Z}{\mathbb{Z}}
\newcommand{\Ind}{\mathrm{Ind}}

\begin{document}


\title{
Coarse geometric approach to topological phases: \\ Invariants from real-space representations}

\author{Christoph S. Setescak}
\affiliation{%
Institute of Experimental and Applied Physics, University of Regensburg, Universitätstraße 31, 93080 Regensburg, Germany}
\author{Caio Lewenkopf}%
\affiliation{%
Instituto de Física, Universidade Federal Fluminense, 24210-346 Niterói, Rio de Janeiro, Brazil}
\author{Matthias Ludewig}%
\affiliation{%
Faculty of Mathematics, University of Regensburg, Universitätstraße 31, 93080 Regensburg, Germany}


\date{\today}

\begin{abstract}
We show that topological phases include disordered materials if the underlying invariant is interpreted as originating from coarse geometry.
This coarse geometric framework, grounded in physical principles, offers a natural setting for the bulk-boundary correspondence, reproduces physical knowledge, and leads to an efficient and tractable numerical approach for calculating invariants. 
As a showcase, we give a detailed discussion of the framework for  three-dimensional systems with time-reversal symmetry. 
We numerically reproduce the known disorder-free phase diagram of a tunable, effective tight-binding model and analyze the evolution of the topological phase under disorder.
\end{abstract}
\maketitle




\paragraph*{Introduction.--} 
Exploring topological effects in the quantum mechanical description of matter is a new frontier in solid state physics \cite{Hasan2010,Qi2011}.
Research in this area is at the interface of fundamental mathematical concepts \cite{Bellissard1992, DeNittis2015, Thiang2016, Thiang2017, Ludewig2022_cob}, theoretical modelling \cite{Vanderbilt2018, Aguilera2019, Moessner2021}  and experiments \cite{Konig2007, Hsieh2008, Chen2009, Sobota2012, Schmid2021}, and aims to harness the unique properties of topologically originated phenomena for technological applications \cite{Vandenberghe2017,Schumer2022}. 
Whereas previously mainly spectral properties have been the focus of solid state physics in determining the emergent large scale electronic behaviour 
\cite{Martin2004}, these topological phases of matter are a novel approach to the classification of materials. 
A geometric entity can be constructed from the occupied bulk energy bands of a translationally invariant Hamiltonian and characterized by topological properties. 
Due to a phenomenon referred to as bulk-boundary correspondence, a non-trivial topology leads to the emergence of a unique gap-closing boundary mode. 
These boundary modes are remarkably robust and topologically protected from disruption by disorder \cite{FuKaneMele2007, Fu2007_2, Kobayashi2013}. 


Characterizing topological phases by defining and calculating topological invariants via electronic band structure theory falls short when dealing with disorder that breaks translational symmetry.
However, topological phase transitions in non-stoichometric materials such as $(\mathrm{Bi}_{1-x}\mathrm{In}_{x})_2\mathrm{Se}_3$ and $(\mathrm{Bi}_{1-x}\mathrm{Sb}_{x})_2\mathrm{Se}_3$ \cite{Barriga2018, Vanderbilt2013,Heremans2017} and more recently in amorphous materials \cite{Corbae2023,Corbae2023EPL}, have been experimentally observed, demonstrating the need to find a rigorous understanding of topological phases of materials without translational invariance. 
Several approaches have been proposed to characterize topological phases of strongly disordered materials, each with its limitations.
These include using the spin Bott index \cite{Hastings2010}, local markers \cite{Bianco2011}, the supercell method 
\cite{Vanderbilt2013}, the scattering matrix \cite{Brouwer2014} and the bulk-boundary correspondence \cite{Focassio2021}. 
Let us consider the most commonly used ones.
The supercell method calculates the standard topological invariant considering an enlarged unit cell containing some disorder distribution. 
This approach offers the advantage that it interfaces nicely with first-principle calculations \cite{Vanderbilt2013}, but the artificial periodicity does not represent a truly disordered material.
The spin Bott index is a two-dimensional invariant \cite{Hastings2010}. Its generalization to three-dimensional materials was constructed in \cite{Focassio2021} in analogy to \cite{FuKaneMele2007} and was empirically shown to give the same results as the supercell method. 
However, as discussed later these two-dimensional invariants do not exist in the presence of general disorder \cite{Ewert2019}, thus we assume that this approach fails for certain disorder configurations.
Finally, using the bulk-boundary correspondence relies on a diagonalization of the Hamiltonian restricted to a bounded region and inspecting possible boundary modes. While this approach is straightforward, the diagonalization can be very costly, it involves choosing an arbitrary boundary and it is difficult to distinguish trivial from topological boundary modes.

There is a powerful alternative approach that is free from all the above discussed limitations:
For nearly four decades, the mathematical framework of $C^*$-algebras and $K$-theory has been proposed to model disorder in condensed matter physics \cite{Bellissard}. 
This approach offers a distinct advantage over the standard methods that utilize the vector bundle representation of occupied states in the Brillouin zone.
The key benefit of the $C^*$-algebraic framework lies in its ability to directly define and compute topological invariants from operators within a suitable $C^*$-algebra of observables. 
Although this approach relies heavily on functional analytic methods, the geometric picture is retained.
Coordinate functions on the Brillouin zone, essential for conventional methods, become non-commutative under multiplication in this framework. 
This transformation results in a \textit{non-commutative space}, a mathematically rigorous and well-suited extension of the Brillouin zone for systems lacking translational invariance due to disorder.


In this approach, the topological invariants take values in the $K$-theory group $K_0(\mathcal{A})$ of the observable $C^*$-algebra $\mathcal{A}$, which consists of the set of projections $p = p^2$ in matrix algebras over $\mathcal{A}$, up to homotopy and stabilization \cite{SM}.
We follow a new development within the above-mentioned non-commutative approach to solid state physics, which uses observable algebras coming from coarse geometry, designed to study the large-scale structure of spaces \cite{Kubota,Ewert2019,ludewig2023coarse}. 
This framework has a clear physical motivation and provides a natural setting for the bulk-boundary correspondence \cite{Ludewig2022_cob}. 

Mathematically, the $K$-theory groups for observable algebras coming from coarse geometry are well understood, but are still abstract objects.
One of our contributions is to demonstrate how to extract numerical information from the abstract $K$-theory classes to determine if a given system is in the non-trivial topological phase.
Explicitly, in this Letter, we implement a concrete algorithm to identify the non-trivial topological nature of a system lacking translational invariance  following the protocol outlined in Ref.~\cite{Doll2021}. 

The results presented here are general and can be accommodated for every Cartan-Altland-Zirnbauer class \cite{Altland1997, Ryu2010, Chiu2016}, but are particularly suited for three-dimensional (3D) topological insulators (TIs) with time reversal symmetry, for which an alternative real-space approach is lacking. Furthermore, 3D systems exhibit a plethora of disorder in experimental practice, thus further motivating the application of our method to 3D systems \cite{Corbae2023, Liebig2022}. 
To demonstrate the method's functionality and its numerical efficiency, we apply it to a well-established low-energy tight-binding model of a 3DTI \cite{Leung2012}.
We discuss the topological phase diagram of the pristine material and analyze the evolution of the topological phase as a function of disorder strength.

\paragraph*{Framework.--} 
We use an effective tight-binding like approach, using the complex Hilbert space $\mathcal{H} = \ell^2(X) \otimes K$, where $X$ is the discrete metric space of the lattice sites and $K$ is the Hilbert space of internal degrees of freedom.
For example, we may take $X \cong \mathbb{Z}^d \subset \mathbb{R}^d$ in the case of a bulk system and $X \cong \mathbb{Z}^{d-1} \times \mathbb{N} \subset \mathbb{R}^d$ for a half space system. 

Any bounded operator $T$ on $\mathcal{H}$ can be written as a 
set of hopping operators $T_{x,y} = \langle x\vert T \vert y\rangle$ 
indexed by lattice sites $x,y \in \mathbb{R}^d$. 
The $T_{x,y}$ are operators on the Hilbert space $K$.
An operator $T$ has finite propagation $R > 0$ if $T_{x,y} = 0$ whenever $\|x-y\| > R$. 
Furthermore, $T$ is called locally compact if all hopping operators $T_{x,y}$ are compact operators. 
The closure with respect to the operator norm of the set of locally compact, finite propagation operators, denoted by $C^*(X)$, is called the Roe algebra of $X$. We remark here that for mathematical reasons, the coefficient Hilbert space $K$ should be taken infinite-dimensional. 
Physically, this corresponds to the presence of infinitely many conduction bands that are typically present but irrelevant for the description of the system.

The Roe algebra depends only on the coarse structure of the space $X$, which roughly means that spaces sharing the same large scale geometry yield isomorphic algebras~\cite{Ewert2019, Roe1996, Higson2000}.

Most Hamilton operators $H$ considered in physics are elements of $C^*(X)$.
We say that such a Hamiltonian is insulating if the Fermi energy $E_F$ is not contained in its spectrum, 
\begin{equation}
    E_F \notin \mathrm{spec}(H).
\end{equation}
Associated to the presence of the spectral gap at $E_F$, the Fermi projection $p_F$, i.e., the spectral projection onto the spectrum below $E_F$, is an element of the Roe-algebra \cite{ludewig2023coarse}.
Analogous to the translational symmetric case, we expect $p_F$ to represent the topological phase. 
Formally, the topological invariant which determines the phase of $H$ is the class $[p_F] \in K_0(C^*(X))$.


For a generic quantum mechanical systems without internal symmetries, the observable algebra is complex, but in our case, we are interested in time-reversal symmetric phases.
In other words, we assume the presence of a fermionic time-reversal operator $\mathrm{T}$, which is an anti-unitary operator acting on $\mathcal{H}$ that commutes with complex conjugation, $\overline{\mathrm{T}} = \mathrm{T}$, and squares to $\mathrm{T}^2 = -1$. The algebra of time-reversal symmetric observables is the subalgebra of operators commuting with $\mathrm{T}$, 
\begin{equation}
    \label{TRSAlg}
    C^*(X)^{\mathrm{T}} = \{a \in C^*(X) \mid \mathrm{T} a \mathrm{T}^{-1} = a\}.
\end{equation}


\paragraph*{The topological invariant.--}
The algebra $C^*(X)^{\mathrm{T}}$ is a real $C^*$-algebra, and the appropriate group of values for topological invariants is the real $K$-theory group $KO_0(C^*(X)^{\mathrm{T}})$. 
In the case of a 3D lattice system, this group in question may be calculated as
\begin{equation}
\label{GroupIsZ2}
  KO_0(C^*(X)^{\mathrm{T}}) \cong \Z/2.
\end{equation}
In other words, up to equivalence, there is only one non-trivial phase. A feature of the theory is coarse invariance, which has the consequence that this result holds much more generally, for example whenever $X$ is a Delone set in $\R^3$. 
In a mathematically precise way, the non-trivial phase obtained in this framework corresponds precisely to the strong topological phase defined by Bloch theory, the weak invariants are lost in this framework \cite{Ewert2019}.

The most important consequence of a non-trivial topological invariant is the emergence of a gap-closing boundary mode: Let $X$ be any Delone set in $\R^3$ and let $Y \subseteq X$ be some subset.
Let $\widetilde{H}$ be the restriction of the gapped Hamiltonian  $H$ to $Y$ with arbitrary local boundary conditions.
Using $K$-theoretic methods, one may then show that when passing from $H$ to $\widetilde{H}$, the spectral gap at $E_F$ is completely filled \cite{SchuBaKellendonkRichter,LudewigThiang,ludewig2023coarse}.
Since $H$ and $\tilde{H}$ coincide away from the boundary the states filling the spectral gap must be boundary-localized.
To draw this conclusion, we only need the ``wideness'' assumption on $Y$, stating that $Y$ contains points of arbitrary distance from $X \setminus Y$, and the same with $Y$ and $X \setminus Y$ exchanged.
For example, $Y$ could be the  intersection of $X$ with any half space, but much more general subsets satisfy this criterion. 
Furthermore the gap closing boundary mode is delocalized, i.e. it is not localized in any bounded region on the boundary.

Another phenomenon associated to a non-trivial invariant 
in 3D systems is the quantized topological magnetoelectric effect \cite{Wang2015}. 
An electric field induces a magnetic field in the same direction, with the coefficient of proportionality quantized in units of $e^2/2\hbar$, analogous to the quantum Hall effect \cite{Klitzing1980}. 
The relevant invariant is the second Chern class of a homotopy between time-reversal symmetric Hamiltonians and has been described in a non-commutative framework, for restricted disorder configurations described by crossed product algebras \cite{Prodan2013, Leung2013}.

As discussed in the introduction, in systems with translational symmetry, the Fu-Kane formula \cite{FuKane,FuKaneMele2007} offers a mathematically well-defined method to identify non-trivial topology. However, this approach becomes inapplicable when the Hamiltonian $H$ and its corresponding Fermi projection $p_F$ lack translational invariance.

For disordered systems, a way to detect the topological invariant, {\it i.e.}, an explicit description of the identification of Eq.~\eqref{GroupIsZ2}, is given by index theory.
The starting point is the self-adjoint operator
\begin{align}
    D = V + \sum_{\mathbf{x} \in X} \gamma_1 \, \hat{x}_1 + \gamma_2 \, \hat{x}_2 + \gamma_3 \, \hat{x}_3,
\end{align}
acting on the Hilbert space $\hat{\mathcal{H}} = \mathcal{H} \otimes \C^2$.
Here $ \hat{x}_i$ are the position operators, $V =\frac{1}{4}\,\gamma_1\,\delta(\hat{x}-0)$ is a small perturbation localized at the origin, such that $D$ is invertible, and the \{$\gamma_i\}$ are Pauli matrices acting on additional degrees of freedom, distinct from the orbital and spin degrees of freedom, for which we use $\{\sigma_i\}$. 
Let $F = \chi(D > 0)$ be the spectral projection onto the positive spectral subspace of $D$.
Then for any projection $p \in C^*(X)$, we set
\begin{align}
    \label{IP}
    T_p = F(1-2p)F + 1 - F,
\end{align} 
an operator on $\ell^2(X) \otimes K \otimes \C^2$.
One may show that $T_p$ has the property that its $\Z/2$-index
\begin{equation}
\label{Z2Index}
 \Ind_2(T_p) = (-1)^{\dim\ker(T_p)}
\end{equation}
is invariant under small perturbations of $p$, hence is well defined on $K$-theory classes and gives an explicit description of the identification \eqref{GroupIsZ2} \cite[\S3]{GrossmannSchulzBaldes}.
In \eqref{Z2Index}, the two-element group $\Z/2$  is represented multiplicatively as~$\{\pm 1\}$.

\paragraph*{Numerical Methods.--}  

Following \cite{Doll2021}, the $\Z/2$-index can be numerically computed as follows.
For some cut-off radius $\rho>0$, consider the finite volume Hilbert space
\begin{equation}
    \hat{\mathcal{H}}_\rho = \chi(|D| \leq \rho) = \ell^2(X_\rho) \otimes K \otimes \C^2,
\end{equation}
where $X_\rho = \{\mathbf{x} \in X \mid |x_1|+|x_2|+|x_3| \leq \rho\}$, and denote by $H_\rho$ and $D_\rho$ the truncations of $H$ and $D$ to $\mathcal{H}_\rho$.
Then by \cite[\S5.7]{Doll2021}, expanding upon a result of  \cite{Loring2015}, we have
\begin{equation}
    \label{sgn_det}
     \mathrm{Ind}_2(T_p) = \frac{\mathrm{sgn}(\mathrm{det}( i D_{0,\rho} - \kappa(H_\rho - E_F))}{\mathrm{sgn}(\mathrm{det}(iD_{0,\rho}))},
\end{equation}
for a suitable tuning parameter $\kappa$, which is required to satisfy the following sufficient, but not necessarily optimal bounds
\begin{equation}
\begin{aligned}
     \label{bounds}
     \kappa &< 1/12\, \| H\|^{-1}  \| [H,D]\|^{-1} \|(H-E_F)^{-1}\|^3, \\
     \rho &> 2 / \kappa \, \|(H-E_F)^{-1}\| .
\end{aligned}
 \end{equation}
Below, we discuss how to choose $\kappa$ and $\rho$ for specific systems while minimizing numerical demand.
The topological invariant as given by Eq.~\eqref{sgn_det} can be evaluated numerically efficiently using the sparse \texttt{LU}-decomposition \cite{MATLAB}, after which the sign of the determinant can be calculated via the complex phases of the diagonal entries of the resulting triangular matrices
\footnote{Reference~\cite{Loring2015} reports the computation of the topological invariant $\mathrm{Ind}_2(T_p)$, but for finite model systems only. 
}.

\paragraph*{Low-Energy Model and Numerical Results.-- } 
\begin{figure}
        \centering
        \includegraphics[width=0.85\linewidth]{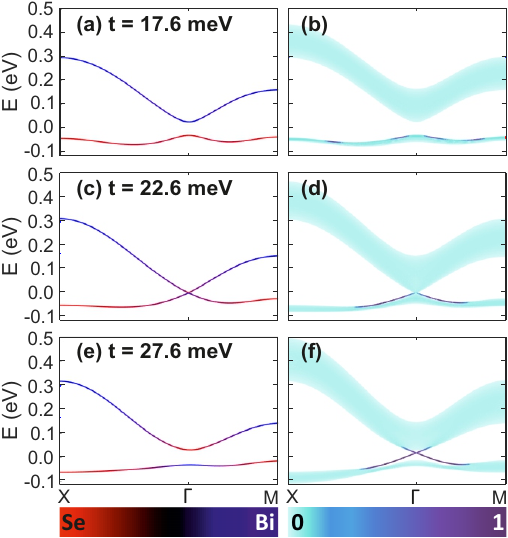}
        \caption{\label{LowEnergy_Bi2Se3}
        Bulk (left) and halfspace (right) bandstructures for different values of $t$. The color scales stand for the intensity of the orbital character (left) and   the projection into the material boundary (right).
        (a) Trivial phase with non-inverted band gap. (b) Upon restricting the material to the halfspace the band gap remains open. At the phase transition the bulk (c) and the halfspace (d) gaps close. (e) In the non-trivial phase the band gap is inverted. (f) The boundary mode emerges upon restricting the system to the halfspace.}
\end{figure}
As a showcase, we use our method to analyze the prototypical model Hamiltonians for 3DTIs on $X = \Z^3$  introduced in \cite{Zhang2009}. 
An application of the method to the Kane-Mele model, which is analogous for the most part, is given in the SM~\cite{Kane2005, Kane2005_2, Loring2017, Wimmer2012, SM}.
We denote by $\hat{S}_{i,j,k}$ the shift operator on $\mathbb{Z}^3$ with shift $i \mathbf{a}_1+j \mathbf{a}_2+k\mathbf{a}_3$, an integer multiple of the lattice vectors $\mathbf{a}_1, \mathbf{a}_2$ and $\mathbf{a}_3$. Expressed in terms of the shift operators the Hamiltonian acting on $\mathcal{H} = \ell^2(\mathbb{Z}^3, \mathbb{C}^4)$ is given as:
\begin{equation}
    \begin{split}
        H_t =\, & \mathrm{id}_{\Z^3} \otimes (\epsilon \, \sigma_0 \otimes \sigma_3 + 6 \gamma) \\
        -\, & i\lambda \sum_{s = \pm 1} s\, \hat{S}_{0s0}\otimes\sigma_2\otimes \sigma_1\\
        +\, &i\lambda \sum_{s = \pm 1} s\,( \hat{S}_{s00}\otimes \sigma_1 \otimes \sigma_1+\hat{S}_{00s}\otimes\sigma_3 \otimes \sigma_1)\\
        -\,&\sum_{s = \pm 1}(\hat{S}_{s00}+\hat{S}_{0s0}+\hat{S}_{00s})\otimes(t \, \sigma_0 \otimes \sigma_3 +\gamma),
    \end{split}
\end{equation}
where the basis of the internal Hilbert space $\C^4$ is $\{\vert \mathsf{P}_{\rm Se},\uparrow \rangle,\vert \mathsf{P}_{\rm Bi},\uparrow \rangle,\vert \mathsf{P}_{\rm Se},\downarrow \rangle,\vert \mathsf{P}_{\rm Bi},\downarrow \rangle \}$.
The parameters are $\epsilon = \SI{134}{meV}$, $\lambda = \SI{30}{meV}$, $\gamma = \SI{16}{meV}$, and $t \in [\SI{14}{meV},\SI{40}{meV}]$, as given in \cite{Leung2012}. It has been shown that for $t < \SI[locale=US]{22.6}{meV}$ the system is in the trivial phase while for $t > \SI[locale=US]{22.6}{meV}$ it becomes non-trivial \cite{Leung2012}. 

Fig.~\ref{LowEnergy_Bi2Se3} shows the bulk and halfspace band structures for the system along a trivial to topologically non-trivial phase transition.
The bulk spectral properties indicate the non-trivial electronic properties, evident from the band inversion observed in Fig.~\ref{LowEnergy_Bi2Se3}(c) compared to (a), and the closure of the bulk band gap occurring at the critical hopping parameter $t_{\rm crit} = \SI[locale=US]{22.6}{meV}$. 
This non-trivial electronic structure leads to the emergence of the gap-closing boundary mode, displayed in Fig.~\ref{LowEnergy_Bi2Se3}(f), whereas (d) the halfspace system for $t = \SI[locale=US]{17.6}{meV}$ remains insulating.

\begin{figure}[h]
        \centering
        \includegraphics[width=0.9\columnwidth]{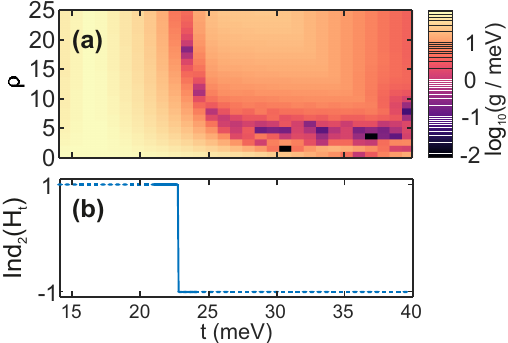}
        \caption{\label{PhaseDiagramCombined} (a) Contour plot showing the spectral gap $g$ of the localizer $L_{1,\rho}$ as a function of $\rho$ and $t$. For all values of $t$, except at the phase transition, the spectral gap $g$ converges to a non-zero value for increasing $\rho$.  (b) Topological invariant as a function of $t$ for $\rho = 30$. The phase transition occurs at $t_{\rm crit} = \SI[locale=US]{22.6}{meV}$.} 
\end{figure}

Constructing the shift operators with Dirichlet boundary conditions on  $\Z^3_\rho$ is achieved by ordering ${\bf x}_i \in \Z^3_\rho$. Then a matrix representation of $\smash{\hat{S}}_{a,b,c,\rho}$ is given by $(\smash{\hat{S}}_{a,b,c,\rho})_{ij} = \delta({\bf x}_i-{\bf x}_j, a {\bf a}_1+b {\bf a}_2+c {\bf a}_3)$. 
In \texttt{Matlab} this approach can be optimized by employing array operations and using logical array indexing, facilitating an efficient and adaptable numerical execution \cite{MATLAB}.

Let us discuss the bounds, Eq.~\eqref{bounds}, in the example of $t = \SI[locale=US]{27.6}{meV}$. 
We calculate $\|H\| = \SI[locale=US]{491}{meV}$ and $g = \SI[locale=US]{14.8}{meV}$ \cite{MATLAB}. As the onsite terms of $H$ commute with $D$ we get $\|[H,D]\| \leq 6(\lambda+t+\gamma) = \SI[locale=US]{442}{meV}$ by making use of the triangle inequality. 
This results in $\kappa < \SI{1.2}{\micro eV}$ and thus the cut-off radius in units of the lattice parameter is $\rho > 2373645$, which is larger than any computationally feasible $\rho$ by about three orders of magnitude. 
But, as neither the bounds \eqref{bounds} nor this estimate are optimal, we revert to a different method for demonstrating convergence \cite{Doll2021}. 
The topological invariant is well defined if the self-adjoint \textit{spectral localizer} $L_{\kappa, \rho} = \kappa D_\rho \otimes \sigma_1 - H_\rho \otimes \sigma_3$ has a finite spectral gap $\widehat{g}$ that approaches a fixed value for $\rho \rightarrow \infty$. Numerically we calculate $\widehat{g}$ with a Krylov-Schur algorithm \cite{MATLAB, Stewart2001}. The result of this test is shown in Fig.~\ref{PhaseDiagramCombined}(a). 
In the trivial phase, the gap of the localizer $L_{1,\rho}$ is greater than zero for all values of $\rho$. In the non-trivial case the gap of the localizer $L_{1,\rho}$ closes around $\rho = 6$ and then opens again for increasing $\rho$. Convergence is slowest around the phase transition. The phase diagram calculated with $\rho = 30$ is shown in Fig.~\ref{PhaseDiagramCombined}(b).
The resolution of the calculation around the phase transition is $10 \, \mathrm{points}/{\mathrm{meV}}$. 
We find $t_{\rm crit} = \SI[locale=US]{22.6}{meV}$, which precisely reproduces the result of \cite{Leung2012}. 

\begin{figure}[h]
        \centering
        \includegraphics[width=0.9\columnwidth]{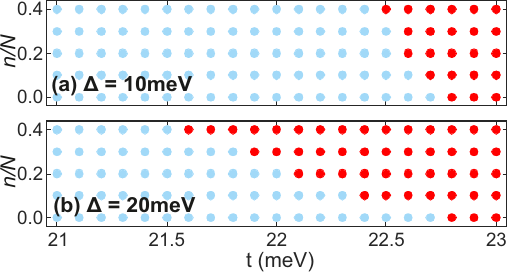}
        \caption{\label{PhaseDiagramDisorder}
        Evolution of the topological phase diagram 
        of a system described by $H_t^{\rm dis}$
        as function of the disorder concentration $n/N$ 
        and $t$ for the disorder amplitudes (a) $\Delta = \SI{10}{meV}$ and (b) $\Delta = \SI{20}{meV}$.
        Red dots indicate the non-trivial phase and light blue dots 
        the trivial one.}
\end{figure}

An extension of this result to Hamiltonians lacking translational symmetry is shown in Fig.~\ref{PhaseDiagramDisorder}.
To break the translational symmetry an onsite potential of $\Delta = \pm  \SI{10}{meV}, \pm \SI{20}{meV}$ was added to a 
randomly chosen subset of  lattice sites $I_\pm$, with $|I_+| = |I_-|= n$
The disordered Hamiltonian $H_t^{\rm dis}$ reads
\begin{equation}
H_t^{\rm dis} = H_t + \Delta \sum_{i \in I_+} \delta(\hat{\bold{x}}-\bold{x}_i) - \Delta \sum_{i \in I_-} \delta(\hat{\bold{x}}-\bold{x}_i).
\end{equation}
The critical hopping parameter $t_{\mathrm{crit}}$ decreases for an increasing perturbation amplitude $\Delta$ and an increasing disorder concentration, $n/N$, where $N$ is the number of lattice points in $X_\rho$. 
In Fig.~\ref{PhaseDiagramDisorder}(a) the result is shown for $\Delta = \SI{10}{meV}$, where the effect is less pronounced than for $\Delta = \SI{20}{meV}$ shown in Fig.~\ref{PhaseDiagramDisorder}(b). Previously, a similar increase of the non-trivial phase upon introduction of disorder for the 2D Kane-Mele-model has been predicted, where non-triviality was demonstrated through quantized conductance \cite{Orth2016}. As quantized conductance does not occur in 3DTIs and other methods used in 2DTIs 
\cite{Assuncao2024} are not applicable in 3D, the new method presented here is necessary, which facilitates direct computation of the topological invariant for the disordered system.

\paragraph*{Conclusion and outlook.--} 
The numerical evidence presented here and the mathematical rigorous interpretation of many effects previously understood in a physical context provide strong arguments for extending the notion of topological phases to materials with disorder using coarse geometry.

As an immediate application, tunable solid solutions, such as $(\mathrm{Bi}_{1-x}\mathrm{In}_{x})_2\mathrm{Se}_3$ and $(\mathrm{Bi}_{1-x}\mathrm{Sb}_{x})_2\mathrm{Se}_3$ \cite{Vanderbilt2013, Barriga2018, Heremans2017}, and amorphous TIs \cite{Corbae2023, Corbae2023EPL} are ideal platforms to experimentally verify the numerical predictions and further establish that coarse geometry is well suited for describing topological phases. 
Our findings have implications beyond condensed matter and materials science, extending to fields such as photonics \cite{Stutzer2019}, acoustics \cite{Nejad2020}, as well as to other  metamaterials \cite{Mitchell2018}, which are increasingly exploring disordered topological phases.

Based on the straightforward numerical implementation of our approach in tight-binding models, we anticipate that extending to \emph{first-principle} electronic structure methods is highly feasible, which will open new research paths for a realistic description of topological systems.
In addition to the conventional TIs represented by the Cartan-Altland-Zirnbauer classes, new concepts like higher-order TIs have come into focus, and we aim to adapt coarse geometric methods to these alternative topological phases.

\begin{acknowledgments}
\paragraph*{Acknowledgements.--} 
CS thanks Franz J. Giessibl for providing the opportunity to pursue the experimental goals laid out in the present article.
We are deeply indebted to Nora Doll and Hermann Schulz-Baldes for introducing us to the finite volume calculation of $K$-theory invariants.
We are grateful to Magdalena Marganska and Klaus Richter for insightful discussions.
We thank Rupert Huber and Tobias Inzenhofer for providing access to computational hardware used for a subset of the calculations presented.
We acknowledge SFB 1085 (project no.\ 224262486) and SFB 1277 (project no.\ 314695032) for funding.
CL is supported by the Brazilian funding agencies FAPERJ and CNPq. 
\end{acknowledgments}

%

\end{document}


\title{Supplemental Material to 
\texorpdfstring{\\}{}
``Coarse geometric approach to topological phases: \\ Invariants from real-space representations"}

\author{Christoph S. Setescak}
\affiliation{%
Institute of Experimental and Applied Physics, University of Regensburg, Universitätstraße 31, 93080 Regensburg, Germany}
\author{Caio Lewenkopf}%
\affiliation{%
Instituto de Física, Universidade Federal Fluminense, 24210-346 Niterói, Rio de Janeiro, Brazil}
\author{Matthias Ludewig}%
\affiliation{%
Faculty of Mathematics, University of Regensburg, Universitätstraße 31, 93080 Regensburg, Germany}

\date{\today}

\maketitle

\section{Connection to Crystalline Case}
We use an effective tight-binding like approach, using the complex Hilbert space $\mathcal{H} = \ell^2(X) \otimes K$, where $X$ is the discrete metric space of the lattice sites and $K$ is the Hilbert space of internal degrees of freedom. 
For crystalline materials, we may take $X = \Z^d$, the periodic lattice, and the algebra of observables may be replaced by the subalgebra of translational invariant operators $C^*_{\mathrm{inv}}(X) \subset C^*_{\Roe}(X)$. 
An operator $T \in C^*_{\Roe}(X)$ is translational symmetric if the hopping matrices $T_{\mathbf{xy}}$ only depends on $\mathbf{x}-\mathbf{y}$, i.e  $T_{\mathbf{xy}} = T(\mathbf{x}-\mathbf{y})$. 
Let $C^*_{\mathrm{group}}(\Z^d)$ be the group $C^*$-algebra of $X$, then 
\begin{align}
    C^*_{\mathrm{inv}}(X) \cong C^*_{\mathrm{group}}(X) \otimes \mathbb{B}(K) \cong C(T^d,\mathbb{B}(K)).
\end{align}
The second isomorphism is the Bloch theorem or Fourier transform. The Fourier transform of the Hamiltonian $H$ is a map $H : \mathbb{T}^3 \rightarrow \mathbb{B}(K)$ from the Brillouin torus $\mathbb{T}^3$ into the space $\mathbb{B}(K)$ of bounded operators on $K$. 
If $H$ is an insulator then $\mathrm{spec}(H)$ has a gap around $E_F$. 
This implies that $\mathrm{spec}(\hat{H}(k))$ is also gapped at $E_F$ for all $k \in \mathbb{T}^3$. Therefore the vector spaces 
\begin{equation}
    \mathcal{V}_k = \bigoplus_{\lambda < E_F} \mathrm{Eig}(H(k),\lambda)
\end{equation}
form a smooth vector bundle $\mathcal{V}$ over the Brillouin zone $\mathbb{T}^3$. 
This bundle is referred to as the Bloch bundle. 
As discussed in the main text, the algebra of observables for time reversal symmetric materials is $\mathcal{A} = C^*(X)^{\mathrm{T}}$. 
After a Fourier transform in the crystalline case, a time reversal symmetric Hamiltonian obeys $\mathrm{T} H(k) \mathrm{T}^* = H(\tau(k))$ with the involution $\tau: \mathbb{T}^3 \rightarrow \mathbb{T}^3$ being given by $\tau(k) =  -k$, such that $\tau^2 = \mathrm{id}$. 

This fermionic time-reversal symmetry gives the Bloch bundle $\mathcal{V}$ a quaternionic structure~\cite{Thiang2017}, \emph{i.e.}, $(\mathcal{V}, \mathrm{T}) \in \mathrm{Vec}^{2m}_{\mathfrak{Q}}(\mathbb{T}^3,\tau)$. 
Such quaternionic vector bundles can be classified by the FKMM invariant \cite[Remark~4.5]{ DeNittis2015}:
\begin{equation}
    \kappa: \mathrm{Vec}^{2m}_{\mathfrak{Q}}(\mathbb{T}^3,\tau) \longrightarrow \Z/2 \oplus (\Z / 2)^{\oplus 3},
\end{equation} 
which is a quadruple $\kappa = (\kappa_0,\kappa_1,\kappa_2,\kappa_3)$.
The first summand $\kappa_0 \in \Z/2$ is referred to as the \emph{strong invariant} in the terminology of Ref.\ \cite{FuKaneMele2007} and is a genuine three-dimensional property of the system. 
The invariants $(\kappa_1,\kappa_2,\kappa_3) \in (\Z / 2)^{\oplus 3}$ are the \emph{weak-invariants}, which correspond to \mbox{$\tau$-equivariant} embeddings of two-dimensional tori.


In Ref.\ \cite{Ewert2019}, the map on $K$-theory induced by the the embedding $f: C^*_{\mathrm{inv}}(X) \hookrightarrow C^*_{\mathrm{Roe}}(X)$ was studied. 
It was found that 
\begin{equation}
    f^*: K_0(C^*_{\mathrm{inv}}(X)) \longrightarrow K_0(C^*_{\mathrm{Roe}}(X))
\end{equation}
is split-surjective, with the kernel consisting of the weak topological phases, and its image given by the strong topological phases.
The fact that the $K$-theory classes corresponding to weak topological phases are in the kernel of $f^*$ confirms the conjecture of Fu, Kane and Mele that these weak topological phases are not robust against disorder \cite{FuKaneMele2007}.

\section{Roe Algebra $K$-Theory}
\label{K(Roe)}
In this section, we review the definition of $K$-theory in the case of the Roe algebra, which allows for a simplified description.

Let $X$ be a discrete metric space and let $C^*_{\Roe}(X)$ be the corresponding Roe algebra, i.e., the norm-closure of all bounded operators on the Hilbert space $\mathcal{H} = \ell^2(X) \otimes K$ that are \emph{locally compact} and have \emph{finite propagation}, meaning that each $T_{\mathbf{x}\mathbf{y}}$ is a compact operator on $K$, and there exists $r>0$ such that $T_{\mathbf{x}\mathbf{y}} = 0$ whenever $\mathrm{dist}(\mathbf{x}, \mathbf{y}) \geq r$.

To define the group $K_0(C^*_{\Roe}(X))$, one considers orthogonal projection operators $p$ (\emph{projections} for short) onto closed subspaces $V$ of $\mathcal{H}$ with the property that $p \in C^*_{\Roe}(X)$.
$V$ is called the \emph{range} of $p$.
If $p$ and $q$ are two projections in $C^*(X)$, a \emph{homotopy} between them is a norm-continuous path $(p_t)_{t \in [0, 1]}$ of elements in $C^*(X)$ such that $p_0 = p$ and $p_1 = q$, and such that each $p_t$ is a projection.
For any projection $p$, we denote by $[p]$ the corresponding homotopy class.

If $p$ and $q$ are two projections in $C^*(X)$ onto orthogonal subspaces $V \perp W$ of $\mathcal{H}$, then $p + q$ is the projection onto the direct sum of $V + W \subseteq \mathcal{H}$, hence a projection again.
Any pair of projections $p$, $q$ inside the Roe algebra may be homotoped into a new pair $p^\prime$, $q^\prime$ whose ranges are orthogonal.
On the set of homotopy classes of projections, there is therefore a well-defined addition operation, given by
\begin{equation}
\label{AdditionOperation}
  [p] + [q] = [p^\prime + q^\prime],
\end{equation}
where $p^\prime$ and $q^\prime$ are perturbations of $p$ and $q$ with orthogonal ranges.

The above addition operation gives the set of homotopy classes of projections in $C^*_{\Roe}(X)$ the structure of a commutative semigroup, with the zero projection as neutral element.
The $K$-theory group of $C^*_{\Roe}(X)$ is then given by adding formal inverses to this semigroup, similar to how one obtains the integers from the natural numbers.
Explicitly, one has
\begin{equation*}
    K_0(C^*_{\Roe}(X)) = \big\{[p]-[q] \mid p, q \in C^*_{\Roe}(X)~\text{projections}\big\},
\end{equation*}
where the relation is
\begin{equation}
\label{EquivalenceRelation}
\begin{aligned}
  &[p] - [q] = [\tilde{p}] - [\tilde{q}] \\
  &\qquad \Longleftrightarrow \text{there exists a projection}~r \in C^*_{\Roe}(X)\\
  &\qquad\qquad~ \text{such that}~[p] + [\tilde{q}] + [r] = [\tilde{p}] + [q] + [r].
\end{aligned}
\end{equation}
Here the addition operation \eqref{AdditionOperation} is used.

\begin{rem}
  The above presentation is very similar to the presentation of the integers as
  \begin{equation*}
  \Z = \{ m - n \mid m, n \in \mathbb{N}\},
\end{equation*}
where $m-n$ and $k-l$ represent the same element if and only if $m+l = k+n$.
However, in contrast to the case of the integers, one needs an additional projection $r$ in the equivalence relation \eqref{EquivalenceRelation},  due to the fact that for semigroup of homotopy classes of projections, the equality $[p] + [r] = [q] + [r]$ does not necessarily imply that $[p] = [q]$.
One says that this semigroup violates the \emph{cancellation property}(which, in contrast, is clearly satisfied in $\N$).
For example, if $p$ is any finite rank projection (these are always contained in the Roe algebra) and $r$ is any projection onto an infinite-dimensional subspace $V \subseteq \mathcal{H}$, then $p + r$ is homotopic to $r$. 
In particular, in $K_0(C^*_{\Roe}(X)$, one has $[p] = 0$ for any finite rank projection if $C^*_{\Roe}(X)$ contains an infinite rank projection (which is always the case unless $X$ has only finitely many points).
\end{rem}

\begin{rem}
To define the $K$-theory of general $C^*$-algebras $\mathcal{A}$, one needs to consider projections in matrix algebras over the algebra $\mathcal{A}^+$ obtained from $\mathcal{A}$ by adding a unit, together with an additional relation that identifies projections in different matrix algebras.
    The simplified description given here is possible because $C^*_{\Roe}(X)$ has the special properties of being \emph{quasi-stable} and admitting an approximate unit of projections.
\end{rem}

Apart from $n=0$, there exist groups $K_n(C^*_{\Roe}(X)$ for any $n \in \N$. 
\emph{Bott periodicity} for complex $K$-theory states that
\begin{equation*}
    K_n(C^*_{\Roe}(X)) \cong K_{n+2}(C^*_{\Roe}(X)),
\end{equation*}
so the only other group one has to know is the group $K_{1}(C^*_{\Roe}(X)$, which is somewhat simpler to define than $K_0(C^*_{\Roe}(X)$.
Denote by $C^*_{\Roe}(X)^+ \subset \mathrm{B}(\mathcal{H})$ the unitization of $C^*_{\Roe}(X)$, which is the algebra of operators of the form $T + \lambda I$, where $T \in C^*_{\Roe}(X)$, $\lambda \in \C$ and $I$ is the identity operator of $\mathcal{H}$ (notice that $I \notin C^*_{\Roe}(X)$ due to the local compactness requirement).
We then have
\begin{equation*}
    K_{1}(C^*_{\Roe}(X)) = \{ [u] \mid u~\text{unitary in}~ C^*_{\Roe}(X)^+\},
\end{equation*}
the set of homotopy classes of unitary operators in $C^*_{\Roe}(X)^+$.
Here $u$ and $v$ are homotopic if there exists a norm-continuous path $(u_t)_{t \in [0, 1]}$ of unitaries $u_t$ in $C^*_{\Roe}(X)$ such that $u_0 = u$, $u_1 = v$.
The set $K_{1}(C^*_{\Roe}(X))$ obtains a group structure from composition of operators, $[u]\cdot [v] = [uv]$, which turns out to be commutative.

It is standard that  $K$-theory of the complex Roe algebra for $X = \Z^d$ is given by
\begin{equation}
\label{KtheoryRn}
    K_n(C^*_{\Roe}(X)) = \begin{cases} \Z &\text{for}~n-d = 0 \mod 2\\
    0 &\text{for}~n-d = 1 \mod 2.
    \end{cases}
\end{equation}

\medskip

For the present article, we need $K$-theory for \emph{real} $C^*$-algebras.
The corresponding groups are denoted by $KO_n$, and Bott periodicity in this case states that
\begin{equation*}
    KO_n(C^*_{\Roe}(X)) \cong KO_{n+8}(C^*_{\Roe}(X)).
\end{equation*}
To explicitly describe these groups, we assume that the Hilbert space $\mathcal{H}$ is equipped with an anti-unitary operator $\mathrm{C}$ such that $\mathrm{C}^2 = I$ and which has propagation zero (i.e., is the identity on the $\ell^2(X)$ factor).
We then define the groups $KO_0(C^*_{\Roe}(X))$ and $KO_{1}(C^*_{\Roe}(X))$ as above, but using homotopy classes of projections and unitaries which are real, i.e., commute with $\mathrm{C}$.
(Of course, all homotopies have to go through real elements.)

Similarly, the groups $KO_4(C^*_{\Roe}(X))$ and $KO_3(C^*_{\Roe}(X))$ are defined using projections and unitaries that are \emph{time-reversal symmetric}, i.e., commute with some fixed anti-unitary operator $\mathrm{T}$ such that $\mathrm{T}^2 = -I$.
In the main article, the identification
\begin{equation*}
    KO_0(C^*_{\Roe}(X)^{\mathrm{T}}) = KO_4(C^*_{\Roe}(X))
\end{equation*}
is used.

The remaining $KO$-theory groups can be defined similarly as groups of unitaries or projections with certain extra symmetries, but we do not explicitly consider observable algebras with other than time-reversal symmetry in this work.

For $X = \Z^d$, the $KO$-groups are given as \cite{Ewert2019}
\begin{align*}
    KO_n(C^*_{\Roe}(\mathbb{Z}^d)) \cong 
    \begin{cases}
    \mathbb{Z}    &  n-d = 0 ~ \text{or}~  4 \mod 8\\
    \mathbb{Z}/2  &  n-d = 1 ~ \text{or}~  2  \mod 8\\
    0             & \text{otherwise.}
    \end{cases}
\end{align*}

\section{Bulk-Boundary Correspondence} 


In the setting of coarse geometry, given a subset $Y \subseteq X$, one considers the \emph{localized Roe algebra} $C^*_{\Roe}(Y \subseteq X)$, which is the norm closure of those operators $T \in C^*_{\Roe}(X)$ that are \emph{supported near} $Y$, meaning there exists $r>0$ such that $T_{\mathbf{x}\mathbf{y}} = 0$  whenever one of $\mathbf{x}$ or $\mathbf{y}$ has distance more than $r$ from $Y$.
Clearly, replacing $Y$ be any fattening $Y_r = \{ \mathbf{x} \in X \mid \mathrm{dist}(\mathbf{x}, Y) \leq r\}$ does not change $C^*_{\Roe}(Y \subseteq X)$.
Similarly, we may consider the Roe algebra $C^*_{\Roe}(\partial Y)$, which is the norm closure in $C^*_{\Roe}(Y)$ of those operators that are supported on $Y \cap (X \setminus Y)_r$ for some $r>0$.

The prototypical example we have in mind is 
\begin{equation}
\label{ProtoEx}
    X = \Z^3, \quad Y = \Z^2 \times \N, \quad \partial Y = \Z^2 \times \{0\},
\end{equation}
but coarse invariance of the theory allows to change $X$ and $Y$ quite significantly without changing the result.

The localized Roe algebra $C^*_{\Roe}(\partial Y)$ is a closed ideal in $C^*_{\Roe}(Y)$, hence the quotient $C^*(Y)/C^*(\partial Y))$ is again a $C^*$-algebra.
Notice moreover that we have a $*$-homomorphism
\begin{equation}
\label{QuotientHomomorphism}
    q : C^*_{\Roe}(X) \longrightarrow C^*(Y)/C^*(\partial Y))
\end{equation}
given by just restricting operators on $X$ to $Y$, which is multiplicative modulo operators supported near $\partial Y$.

It then follows from the general theory of $C^*$-algebra $K$-theory that there is an exact sequence
\begin{equation*}
\begin{tikzcd}
    KO_4\big(C^*(Y)\big) \ar[r] & KO_4\big(C^*(Y)/C^*(\partial Y)\big) \ar[d, "\Exp"] \\
    &  KO_{3}\big(C^*(\partial Y)\big),
\end{tikzcd}
\end{equation*}
meaning that the kernel of the vertical map $\Exp$ equals the image of the vertical.
The composition of $\Exp$ with the map on $KO$-theory induced by \eqref{QuotientHomomorphism} yields a group homomorphism
\begin{equation}
\label{MayerVietorisMap}
    \delta : KO_4\big(C^*_{\Roe}(X)\big) \longrightarrow KO_3\big(C^*_{\Roe}(\partial Y)\big).
\end{equation}

An explicit description of $\Exp$ is the following: A class $[p]$  in $C^*_{\Roe}(Y)/C^*_{\Roe}(\partial Y)$ may be represented by a self-adjoint operator $\tilde{p}$ in $C^*_{\Roe}(Y)$ such that $\tilde{p}^2 - \tilde{p} \in C^*_{\Roe}(\partial Y)$, but it is generally not possible to choose $\tilde{p}$ to be a projection in $C^*_{\Roe}(Y)$.
The map $\Exp$ is then defined as
\begin{equation*}
    \Exp([p]) = [\exp(2 \pi i \tilde{p})].
\end{equation*}
Notice that if $\tilde{p}$ is a projection, then $\exp(2 \pi i \tilde{p}) = 1$, which represents the trivial element of $KO_3(C{\rm Roe}^*(Y))$.

Suppose now that $H$ is a gapped time-reversal symmetric Hamiltonian $H$ on $X$, whose Fermi projection $[p_F]$ defines a class in $KO_0(C_{\rm Roe}^*(X)^{\mathrm{T}}) = KO_4(C_{\rm Roe}^*(X))$.
Let $\tilde{H}$ be the Hamiltonian restricted to $Y$, endowed with arbitrary local boundary conditions (meaning that the difference $\tilde{H} - H|_Y$ is supported near $\partial Y$).
One then has the following gap-filling result using the boundary map~\eqref{MayerVietorisMap}:

\begin{theo}
    Suppose that $\delta([p_F])$ is a non-trivial class in $KO_3(C^*_{\Roe}(\partial Y))$.
    Then $\tilde{H}$ is no longer gapped at the Fermi energy of $H$.
\end{theo}

Indeed, if $\tilde{H}$ is gapped at the Fermi energy of $H$, then the corresponding Fermi projection $\tilde{p}_F$ is a lift of the quotient projection $q(p_F) \in C^*_{\Roe}(Y)/C^*_{\Roe}(\partial Y)$ to $C^*_{\Roe}(Y)$, rendering $\delta([p_F]) = \Exp([q(p_F)])$ to be trivial, as $\Exp$ measures the failure of $\tilde{p}_F$ to be a projection.

In the prototypical example \eqref{ProtoEx}, the map $\delta$ is in fact an isomorphism, and one obtains that whenever the Fermi projection $p_F$ of a Hamiltonian $H$ is $KO$-theoretically non-trivial, the restricted Hamiltonian $\tilde{Y}$ has no spectral gap, giving rise to boundary localized states around the Fermi energy. 

In fact, one may also show that while these states are boundary localized, they are delocalized \emph{along} the boundary, i.e., they are not localized to any proper subregion of $\partial Y$.

\section{Numerical Calculation of $KO$-Theory Invariants}
The Fermi projection $p_F \in C^*_{Roe}(\mathbb{Z}^3)$ represents an element $[p_F]$ in the abstract group $KO_4(C^*_{Roe}(\mathbb{Z}^3)) \cong \mathbb{Z}/2$. Determining whether the Fermi projection represents the trivial element or non-trivial element in $\mathbb{Z}/2$ is not straightforward. 
The strategy is to first map $p_F$ to a Fredholm operator $T$ and then numerically evaluate the $\mathbb{Z}/2$-index of this operator, which is based on a version of spectral flow. 
Following Ref. \cite{Doll2021}, this can be done by evaluating the Pfaffian or determinant of a large matrix constructed from the Dirichlet restriction of the Hamiltonian operator on a finite volume.

We shift the Hamiltonian in such a way that the Fermi energy $E_F$ is at zero.
Since $H$ is time-reversal symmetric, so is the Fermi projection $p_F = \chi(H < 0)$.
Denote by $\gamma_i$ the Pauli matrices and consider the self-adjoint operator
\begin{align}
    D_0 = V + \sum_{\mathbf{x} \in X}  \gamma_1 \, \hat{x}_1 + \gamma_2 \, \hat{x}_2 + \gamma_3 \, \hat{x}_2 ,
\end{align}
acting on $\mathcal{H} \otimes \C^2$, and let $E = \chi(D_0 > 0)$ be the corresponding Hardy projection. For $\Sigma = \id_{\mathcal{H}} \otimes i \gamma_2$ the condition $\Sigma^* \overline{D_0}\Sigma = -D_0$ holds. This implies $\Sigma^* \overline{E}\Sigma = \id - E$ (We will suppress the identity $\id_{\mathcal{H}}$ in the following and write $\Sigma = i \gamma_2$). 

The Hamiltonian $H$ acts on $\mathcal{H} \otimes \C^2 \cong \ell^2(\mathbb{Z}^3) \otimes K \otimes \C^2$ by identifying it with $H \otimes \gamma_0$. We write the coefficient Hilbert space $K$ as $K \cong K' \otimes \C^2$, with $\C^2$ being the Hilbert space associated to the spin degrees of freedom. The unitary part of the time-reversal operator $T$ is $\Theta = i \sigma_2$ acting on the factor $\C^2$ associated to the spin degrees of freedom, with the identities on all other factors suppressed in the notation.

We note that, for applying the protocol of \cite{Doll2021} it is necessary for the symmetries $\Theta$ and $\Sigma$ to commute, i.e. $[\Theta,\Sigma] = 0$, which holds true.

In any physical relevant situation we can assume that $H$ is approximated to any arbitrary precision by the restriction to only finitely many electronic degrees of freedom, this amounts to $K \cong \mathbb{C}^{2n}$ for $n \in \N$.  

One may show that the commutator $[p_F, E]$ is compact for any gapped Hamiltonian $H$ in the Roe algebra $C^*_{\Roe}(\Z^3)$, hence the operator 
\begin{align}
    \label{T_3D}
    T = E(1-2p_F)E + 1 - E
\end{align} 
is a self-adjoint Fredholm operator with $\Ind(T) = 0$ \cite{Doll2021}, hence its
        $\mathbb{Z}/2$-index defined by 
        \begin{equation*}
            \mathrm{Ind}_2(T) = (-1)^{\mathrm{dim}(\ker(T))} \in \{\pm 1\}
        \end{equation*} is constant under norm-continuous homotopies of $p_F$ \mbox{and $E$ \cite[Theorem 5]{Doll2021}.}


Following reference \cite{Doll2021} we repeat the construction of the \emph{finite volume skew-localizer} $\smash{\widehat{L}_{\kappa,\rho}}$, which is a finite-dimensional matrix built from $H$ and $D_0$, whose Pfaffian classifies the topological phase. The tuning parameter $\kappa > 0$ needs to be sufficiently small and the parameter $\rho > 0$, which is a measure for the size of the finite volume, needs to be sufficiently large. Precise bounds can be placed on $\kappa$ and $\rho$ and are given as sufficient conditions in Theorem \ref{finiteVolume}. 

The even \emph{spectral localizer} can be constructed from the doubled Hamilton operator $H$ and the operator $D_0$ as
\begin{equation}
L_\kappa = \begin{pmatrix}
            - H & \kappa D_0^*\\
            \kappa D_0 & H
            \end{pmatrix}.
\end{equation}
Due to the the fact that $\Theta$ and $\Sigma$ commute the even \emph{spectral localizer} $L_\kappa$ has the property, that $Q^* L_\kappa Q  = -L_\kappa$. In the three-dimensional, time-reversal symmetric case, $Q$ is defined as
\begin{equation}
Q = \begin{pmatrix}
        0 &  \mathcal{S}\\
        \mathcal{S} & 0
        \end{pmatrix},
\end{equation}
with $\mathcal{S} = \Sigma \, \Theta$. Next we find that the operator $R$ with
\begin{equation}
\label{skewlocalizer}
R = \frac{1}{2}\begin{pmatrix}
        (1-i)\mathcal{R} &  (1+i)\mathcal{R}\\
        (1+i)\mathcal{R} & (1-i)\mathcal{R}
        \end{pmatrix},
\end{equation}
is a unitary root of $Q$, i.e. $\mathcal{R}^2 = \mathcal{S}$ and $\overline{\mathcal{R}} = \mathcal{R}^*$. 
As a consequence the \emph{skew-localizer} 
\begin{equation}
    \label{skew_loc}
    \widehat{L}_\kappa = i R^* L_\kappa R
\end{equation}
is real and skew-symmetric. For the particular dimension and symmetries at hand we calculate explicitly
\begin{equation}
\label{L_k}
\widehat{L}_\kappa = \begin{pmatrix}
        0 &  \mathcal{R}^*(i\kappa D_0 + H)\mathcal{R}\\
        \mathcal{R}^*(i\kappa D_0 - H)\mathcal{R} & 0
        \end{pmatrix}.
\end{equation}


The operator $\widehat{D}_{1}$ is defined analogously by replacing $H$ in Equation \eqref{L_k} with $0$.
The spectral properties of $\widehat{D}_{1}$ are used to describe the restriction to a finite volume. 
We note that $\widehat{D}_{1}$ has a compact resolvent, as the only accumulation point in its spectrum is infinity, thus the Hilbert space
\begin{equation}
    \widehat{\mathcal{H}}_\rho = \mathrm{im(\chi(\vert \widehat{D}_{1} \vert < \rho ))}
\end{equation}
is finite dimensional. Denote by $\widehat{\pi}_\rho = \chi(\vert \widehat{D}_{1} \vert < \rho )$ the partial isometry given by projection onto $\smash{\widehat{\mathcal{H}}_\rho}$.

With this partial isometry, the finite volume restrictions $\smash{\widehat{L}_{\kappa,\rho}}$ and $\smash{\widehat{D}_{1,\rho}}$ are defined as
\begin{equation}
    \begin{aligned}
        \widehat{L}_{\kappa,\rho} = \widehat{\pi}_\rho \, \widehat{L}_\kappa \, \widehat{\pi}_\rho^*
         \quad \mathrm{and} \quad
        \widehat{D}_{1,\rho} = \widehat{\pi}_\rho \, \widehat{D}&_1 \, \widehat{\pi}_\rho^*.
    \end{aligned}
\end{equation} 
Both $\widehat{L}_{\kappa,\rho}$ and $\widehat{D}_{1,\rho}$ are finite matrices. 
Standard functional analytic arguments show that these matrices are merely built from the Dirichlet restrictions of $H$ and $D_0$, i.e., $\smash{\widehat{L}_{\kappa,\rho}}$ and $\smash{\widehat{D}_{1,\rho}}$ are constructed analogously to \eqref{L_k} with $D$ and $H$ replaced by $D_{0,\rho}$ and $H_\rho$. 
These latter matrices are the Dirichlet restrictions of $D_0$ and $H$ to the finite volume
\begin{equation}
    X_\rho = \{\mathbf{x} \in X \mid |x_1|+|x_2|+|x_3| \leq \rho\}.
\end{equation}
The explicit construction of $D_{0,\rho}$ and $H_\rho$ relies on finding a matrix representation of the position and shift operators on $X_\rho$, which is discussed in the main text.
The matrices $\smash{\widehat{L}_{\kappa,\rho}}$ and $\smash{\widehat{D}_{1,\rho}}$ are skew-symmetric and real, thus have a well defined Pfaffian. 
The relevant result of Ref. \cite{Doll2021} is contained in Theorem \ref{finiteVolume}.

\begin{theo}[§31, Ref. \cite{Doll2021}]
    \label{finiteVolume}
    If the bounds on $\rho$ and $\kappa$ given by 
    \begin{align*}
        \begin{aligned}
             \label{bounds}
             \kappa &< 1/12\, \| H\|^{-1}  \| [H,D]\|^{-1} \|(H-E_F)^{-1}\|^3, \\
             \rho &> 2 / \kappa \, \|(H-E_F)^{-1}\| .
        \end{aligned}
    \end{align*}
    are satisfied the $\mathbb{Z}/2$-index can be calculated by the signature of the finite-volume skew-localizer 
    \begin{align*}
        \mathrm{Ind}_2(T)=\mathrm{sgn}(\mathrm{Pf}(\widehat{L}_{\kappa,\rho}))\,\mathrm{sgn}(\mathrm{Pf}(\widehat{D}_{1,\rho})).
    \end{align*} 
\end{theo}

From elementary properties of the Pfaffian of off-diagonal matrices, it follows that if there exists a basis transformation such that the skew-localizer is off-diagonal as in Equation~\ref{L_k}, the formula in the theorem simplifies to
    \begin{equation}
    \label{DeterminantFormula}
        \mathrm{Ind}_2(T) = \mathrm{sgn}(\mathrm{det}( i \kappa D_{0,\rho}-H_\rho)) \,\mathrm{sgn}(\mathrm{det}(iD_{0,\rho})).
    \end{equation}
This simplification is valuable, as efficient algorithms for the calculation of determinants are much more readily available than for the Pfaffian. 

\section{Convergence Analysis}
\label{SM:convergence}
\begin{figure}
        \centering
        \includegraphics{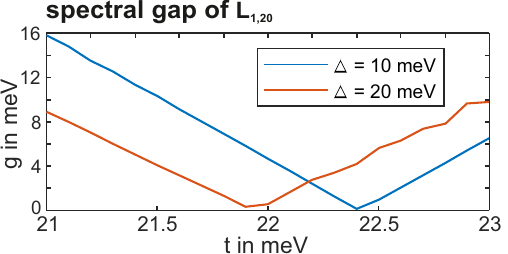}
        \caption{\label{g} Spectral gap of $L_{1,20}$ for disordered three-dimensional topological insulators for $n/N = 0.4$ and a disorder strength of $\Delta = \SI{10}{meV}$ and $\Delta = \SI{20}{meV}$
        (see main text for a detailed model description). 
        At the phase transition the spectral gap closes.}
\end{figure}

As mentioned in the main text, the spectral gap $g$ of the self-adjoint matrix $L_{\kappa,\rho}$ can be used as measure for convergence, as the sign of the Pfaffian cannot change without the spectral gap closing.

Equation~\eqref{skew_loc} holds analogously in the finite volume case, and thus we have $\mathrm{spec} ( \widehat{L}_{\kappa,\rho} ) = i \, \mathrm{spec} \left( L_{\kappa,\rho}\right)$, as the unitary transformation $R_\rho$ leaves the spectrum invariant~\cite{Doll2021}.
Thus, for the sign of the Pfaffian of the skew-localizer $\widehat{L}_{\kappa,\rho}$ to change, the spectral localizer $L_{\kappa,\rho}$ must have an eigenvalue $\lambda$ crossing $0$, implying $g = 0$. 
The second part of our argument relies on the assumption that, if $g$ approaches a finite value for $\rho \gg 1$, as verified numerically, the spectral gap remains open for $\rho \to \infty$.
In other words, stabilization of $g$ for large values of $\rho$ may be taken as an indication of convergence.
However, mathematically, we cannot exclude the possibility of a gap closing between the largest computationally feasible value of $\rho$ and the minimal $\rho$ satisfying the bounds of Thm.~\ref{finiteVolume}.


Figure 2 (in the main text) presents the results of our calculation of the spectral gap $g$ of the spectral localizer $L_{\kappa,\rho}$ for the unperturbed Hamiltonian $H_\rho$ and $\kappa = 1$ as a function of $\rho$, for the model of a three-dimensional topological insulator.
Whenever the parameter $t$ is such that the bulk Hamiltonian $H = H_t$ has a spectral gap, \emph{i.e.}, away from the topological phase transition, the spectral gap of $L_{\kappa,\rho}$ has stabilized for $\rho > 20$.
Furthermore, the obtained $t_{\rm crit}$, which we calculate using the result of Theorem~\ref{finiteVolume}, agrees within three significant digits with the value obtained by a direct inspection of the band dispersion as a function of $t$ (see Figure~1 of the main text).
This conclusively shows that our algorithm has converged for the disorder free case.
As a side note, the lattice parameter of $\mathrm{Bi_2Se_3}$ in the basal plane is \SI{4.14}{\angstrom}. Based on this, a volume with $\rho = 30$ corresponds to a system diameter of approximately \SI{250}{\angstrom}.

For disordered systems, where no direct comparison to a half space bandstructure is available, we proceed similarly.
The data in Figure~\ref{g} corresponds to the top line of the disorder phase diagram in the main text calculated for $\rho = 20$.
Away from the phase transition, the spectral gap $g$ assumes a finite value.
As observed in the pristine system, this indicates convergence of our algorithm in the disordered case.

\section{Kane-Mele Model}
The $K$-theoretic methods introduced in the main text can be applied to any topological phase represented by the Cartan-Altland-Zirnbauer classes \cite{Altland1997}. The $KO$-theory of the relevant Roe algebra $C^*_{\Roe}(\mathbb{Z}^2)$ in two dimensions is given in \ml{\eqref{KtheoryRn}}. Thus the topological invariant of two-dimensional, time-reversal symmetric systems take values in 
\begin{align}
    KO_4(C^*_{\Roe}(\mathbb{Z}^2)) = \mathbb{Z}/2.
\end{align}
The construction of the \emph{skew-localizer} for two-dimensional systems is similar to the three-dimensional case. The major difference is that it is not possible to calculate the Pfaffian via determinants, as the \emph{skew-localizer} in two dimension lacks the symmetries to obtain a transformation to an off-diagonal matrix. As a result the construction of the \emph{skew-localizer} is actually more significant for the numerical implementation.

The operator $D_0$ for two dimensional systems is 
\begin{align}
    D_0 = V + \sum_{\mathbf{x} \in X} \hat{x}_1 + i\,  \hat{x}_2,
\end{align}
where again $V$ is a small perturbation at the origin, such that $D_0$ is invertible \cite{Loring2017}.  Let $F = D_0/|D_0|$ be the phase of the Dirac operator. In two dimensions the symmetry operator $\Sigma = 1$ is trivial. It fulfills the requirements given in \cite{Doll2021}, that
\begin{equation}
    \Sigma^* \, \overline{F}^* \, \Sigma = F \quad \mathrm{and} \quad \Sigma^2 = 1.
\end{equation}
The symmetries of the Hamilton operator $H$ and the Fermi projection $p_F$ are the same as for the three-dimensional case, explicitly 
\begin{equation}
    \Theta^* \, \overline{p_F} \, \Theta = p_F \quad \mathrm{and} \quad \Theta^2 = -1.
\end{equation}
The relevant index pairing, which is analogous to the operator $T$ which we gave for three dimensions in Equation~\ref{T_3D}, is given by 
\begin{equation}
    T = p_F \, F \, p_F + 1 - p_F,
\end{equation}
which is again Fredholm with $\Ind(T) = 0$. The grading operator $Q$ is
\begin{equation}
    Q = \begin{pmatrix}
        0 &  \mathcal{S}\\
        -\mathcal{S} & 0
        \end{pmatrix},
\end{equation}
and the unitary root $R$ of $Q$ is
\begin{equation}
    R = \frac{1+i}{2}\begin{pmatrix}
        \sigma_0 &  \sigma_2\\
        -\sigma_2 & \sigma_0
        \end{pmatrix}.
\end{equation}
With these modifications in place, the construction of the \emph{skew-localizer} is given again by Equation \ref{skewlocalizer}, analogous to the three-dimensional case.
The results of Theorem~\ref{finiteVolume} apply equally, but \eqref{DeterminantFormula} cannot be applied, as $\smash{\widehat{L}_\kappa}$ is not off-diagonal. Evaluating the Pfaffian directly is more involved than for the determinant. 
We use an implementation described in \cite{wimmer2012}, which relies on calculating the Hessenberg form of $\smash{\widehat{L}_\kappa}$. 
The main numerical drawback is that this algorithm is not implemented for sparse matrices in \texttt{Matlab} \cite{MATLAB}.

\begin{figure}[h]
        \centering
        \includegraphics{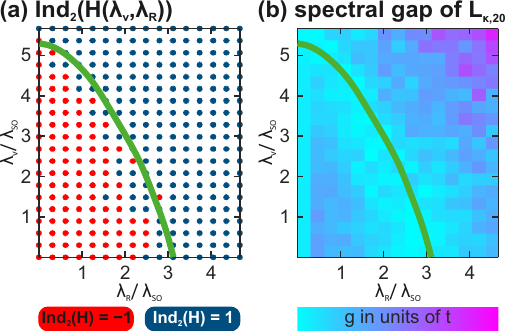}
        \caption{\label{KaneMele}(a) Topological phase diagram of the Kane-Mele  Hamiltonian in the $\lambda_R$ and $\lambda_\nu$ plane. $\mathrm{Ind}_2(H(\lambda_R, \lambda_\nu))$ was calculated with a fixed $\rho = 20$ and variable $\kappa$, i.e. $\kappa(\lambda_R) = 0.005 + \frac{299}{280} \lambda_R$
        (b) Shows the spectral gap of the localizer $L_{\kappa,20}$ in the same $\lambda_R$ and $\lambda_\nu$ plane. It can be seen that the spectral gap of the localizer closes at the phase transition.}
\end{figure}

The Hamiltonian of the Kane-Mele model as a polynomial with entries in the shift operators is \cite{Kane2005_2}:
\begin{equation}
    \begin{split}
        H &= t\sum_{i=1}^{6} \hat{S}_i + i \lambda_{SO} \sum_{i,j=1, i \neq j}^{6} s_z \, \nu_{ij} \, \hat{S}_i\hat{S}_j \\
        &+ i \lambda_{R} \sum_{i=1}^{6}(\bold{s} \times \bold{d}_i)_3\hat{S}_i + \lambda_{\nu} \, (\chi_1 - \chi_2).
    \end{split}
\end{equation}
Here for $i = 1,...,6$ the vectors $\bold{d}_i$ connect nearest neighbour atomic sites (note, these vectors are not the lattice vectors). The shift operator along $\bold{d}_i$ is denoted by $\smash{\hat{S}}_i$. With Dirichlet boundary conditions a matrix representation of the shift operators can be obtained with the same method as for the three dimensional model. 
The first term, which is proportional to the parameter $t$, is a nearest neighbour hopping term. 
The second term is a mirror symmetric spin orbit coupling, given by spin dependent second neighbour hopping. 
It is proportional to the parameter $\lambda_{SO}$ and we have $\nu_{ij} = \smash{\frac{2}{\sqrt{3}}}(\bold{d_i} \times \bold{d}_j)_3 = \pm 1$. In this term $s_z = \sigma_3$ is the $z$-component of the spin operator. The strength of the Rashba spin-orbit-coupling is given by the parameter $\lambda_R$. It is implemented by the operator $(\bold{s} \times \bold{d}_i)_3 = \sum_{n,m = 1}^3 \varepsilon_{n,m,3} \, s_n (\bold{d}_i)_m$. It breaks the mirror symmetry along the $z$-axis. The physical interpretation for this term is given by interaction with a substrate or a perpendicular electric field. The operators $\chi_1$ and $\chi_2$ are the indicator functions on the two sublattices, \emph{i.e.} $\chi_1$ yields 1 for a unit vector placed on atom 1, $0$ if it is placed on atom 2 and vice versa. Multiplied by $\lambda_\nu$ this gives the mass term.  

Fig. \ref{KaneMele} shows the numerical result we obtained for the Kane-Mele model \cite{Kane2005}. For our calculations we used constant $t = 1$ and $\lambda_{SO} = 0.3$. The parameters $\lambda_R \in [0, 1.4]$ and $\lambda_\nu\in [0, 1.7]$ were varied to obtain the phase diagram in Fig. \ref{KaneMele}(a). The boundary of the non-trivial region was graphically extracted from Ref. \cite{Kane2005_2} and overlayed as green line. In Ref. \cite{Kane2005_2} the phase diagram is calculated for $0 < \lambda_{SO} \ll t$, which we assume to be valid for the choice of parameters in our calculation. In whole, we find good agreement between the topological phase diagram given in Ref.~\cite{Kane2005_2} and our numerical result. Discrepancies in the vicinity of the phase diagram will be discussed in the following. The convergence was checked again by calculating the spectral gap of $L_\kappa$ and verifying that it has approached a finite value, which can be seen in Fig.~\ref{KaneMele}(b). As previously demonstrated in three-dimensions, we again notice, that around the phase transition the spectral gap vanishes. In contrast to the three dimensional model, we found that best results were obtained when the parameter $\kappa$ was chosen dependent upon $\lambda_R$, with $\kappa(\lambda_R) = 0.005 + \frac{299}{280} \lambda_R$. As can be seen in the phase diagram displayed in Fig.~\ref{KaneMele}(a) for certain points inside the non-trivial region our method returns a trivial invariant. In general, the drawbacks of using computational tools not optimized for sparse matrices become apparent. We assume, that errors such as the ones in Fig. \ref{KaneMele}(a) would be eliminated for a larger value of the cut-off parameter $\rho$.  


%